\documentclass[fleqn,usenatbib]{mnras}

\usepackage[T1]{fontenc}

\DeclareRobustCommand{\VAN}[3]{#2}
\let\VANthebibliography\thebibliography
\def\thebibliography{\DeclareRobustCommand{\VAN}[3]{##3}\VANthebibliography}

\usepackage{graphicx}	
\usepackage{amsmath}	
\usepackage{amssymb}	
\usepackage{arydshln}   
\usepackage{xcolor}
\usepackage[para,online,flushleft]{threeparttable}

\usepackage{newtxtext,newtxmath}

\title[ICM Metallicities of High Redshift Clusters]{The History of Metal Enrichment Traced by X-ray Observations of High Redshift Galaxy Clusters}

\author[Flores et al.]{
Anthony M. Flores,$^{1,2}$\thanks{E-mail: aflores7@stanford.edu}
Adam B. Mantz,$^{2}$
Steven W. Allen,$^{1,2,3}$
R. Glenn Morris,$^{2,3}$
\newauthor
\ Rebecca E. A. Canning,$^{1,2,4}$
Lindsey E. Bleem,$^{5,6}$
Michael S. Calzadilla,$^{7}$
Benjamin T. Floyd,$^{8}$
\newauthor 
\ Michael McDonald,$^{7}$
Florian Ruppin$^{7}$
\\
$^{1}$Department of Physics, Stanford University, 382 Via Pueblo Mall, Stanford, CA 94305, USA\\
$^{2}$Kavli Institute for Particle Astrophysics and Cosmology, Stanford University, 452 Lomita Mall, Stanford, CA 94305, USA\\
$^{3}$SLAC National Accelerator Laboratory, 2575 Sand Hill Road, Menlo Park, CA 94025, USA\\
$^{4}$ Institute of Cosmology and Gravitation, University of Portsmouth, Burnaby Road, Portsmouth, PO1 3FX, UK\\
$^{5}$HEP Division, Argonne National Laboratory, Argonne, IL 60439, USA\\
$^{6}$Kavli Institute for Cosmological Physics, University of Chicago, 5640 S Ellis Ave, Chicago, IL 60637, USA\\
$^{7}$Kavli Institute for Astrophysics and Space Research, Massachusetts Institute of Technology, Cambridge, MA 02139, USA\\
$^{8}$Department of Physics and Astronomy, University of Missouri, 5110 Rockhill Road, Kansas City, MO 64110, USA
}

\date{Accepted XXX. Received YYY; in original form ZZZ}

\pubyear{2020}

\begin{document}
\label{firstpage}
\pagerange{\pageref{firstpage}--\pageref{lastpage}}
\maketitle

\begin{abstract}
We present the analysis of deep X-ray observations of 10 massive galaxy clusters at redshifts $1.05 < z < 1.71$, with the primary goal of measuring the metallicity of the intracluster medium (ICM) at intermediate radii, to better constrain models of the metal enrichment of the intergalactic medium.  The targets were selected from X-ray and Sunyaev-Zel'dovich (SZ) effect surveys, and observed with both the \textit{XMM-Newton} and \textit{Chandra} satellites. For each cluster, a precise gas mass profile was extracted, from which the value of $r_{500}$ could be estimated. This allows us to define consistent radial ranges over which the metallicity measurements can be compared. In general, the data are of sufficient quality to extract meaningful metallicity measurements in two radial bins, $r<0.3r_{500}$ and $0.3<r/r_{500}<1.0$. For the outer bin, the combined measurement for all ten clusters, $Z/Z_{\odot} = 0.21 \pm 0.09$, represents a substantial improvement in precision over previous results. This measurement is consistent with, but slightly lower than, the average metallicity of 0.315 Solar measured at intermediate-to-large radii in low-redshift clusters. Combining our new high-redshift data with the previous low-redshift results allows us to place the tightest constraints to date on models of the evolution of cluster metallicity at intermediate radii.  Adopting a power law model of the form $Z \propto \left(1+z\right)^\gamma$, we measure a slope $\gamma = -0.5^{+0.4}_{-0.3}$, consistent with the majority of the enrichment of the ICM having occurred at very early times and before massive clusters formed, but leaving open the possibility that some additional enrichment in these regions may have occurred since a redshift of 2.

\end{abstract}




\section{Introduction}
As the most massive gravitationally bound structures in the Universe, the deep gravitational wells of galaxy clusters trap essentially all baryonic matter present during their formation and subsequent evolution \citep*{Allen1103.4829, Kravtsov1205.5556}. Metals produced by stellar processes and ejected from galaxies within these volumes mix with the hot intracluster medium (ICM). X-ray spectroscopic techniques allow us to determine accurate elemental abundances for the ICM \citep{Boehringer2010A26ARv..18..127B, Mernier1811.01967} and, by making measurements across a range of redshifts, construct the histories of star formation and metal enrichment in our Universe.

The metallicity of the ICM in the centers of low redshift clusters is often centrally peaked \citep{Allen9802219,De-Grandi0012232,De-Grandi0310828} and has been shown to evolve moderately with redshift, albeit with substantial intrinsic scatter (e.g.\ \citealt{Mantz1706.01476}) indicative of ongoing and somewhat sporadic enrichment and mixing in these regions. In contrast, the metallicity at intermediate-to-large radii is observed to be remarkably uniform and shows no evidence of evolution. In particular, detailed \textit{Suzaku} observations of the nearest, X-ray brightest galaxy clusters, including the Perseus \citep{Werner1310.7948}, Coma \citep{Simionescu1302.4140}, and Virgo \citep{Simionescu1506.06164} clusters among others \citep{Tholken1603.05255, Urban1706.01567}, found a remarkably uniform distribution of iron, with a metallicity of $Z/Z_{\odot} \sim 0.315\pm0.008$ Solar (combining the independent Suzaku measurements of \citealt{Werner1310.7948} and \citealt{Urban1706.01567}, and using the \citealt{Asplund0909.0948} Solar abundance table). These results extended earlier findings with, in particular, \textit{BeppoSAX} and \textit{XMM-Newton} which determined consistent results at intermediate radii, when scaled to the same Solar abundance table (e.g. \citealt{De-Grandi0012232,De-Grandi0310828, Leccardi0806.1445}; for a recent review see \citealt{Mernier1811.01967}).

Extending to higher redshifts ($z\lesssim 1.3$), measurements of the cluster metallicty at intermediate-to-large radii (e.g. $r\gtrsim 0.3r_{500}$) are challenging to make, due to the low surface brightness and smaller angular scales involved, the latter being prohibitive for instruments like \textit{Suzaku}. Nevertheless, some pioneering studies have been carried out (e.g. \citealt{Ettori1504.02107,McDonald1603.03035,Mantz1706.01476,Liu200312426}) which, to date, have found no significant evidence for evolution.  In particular, a single, very deep \textit{XMM-Newton} measurement of the cluster SPT$-$CL J0459$-$4947 at $z = 1.71$ determined a metallicity consistent with $\sim 0.3$ Solar \citep{Mantz2006.02009}. 

Here, we expand on previous work by filling in the gap in high quality metallicity measurements between the well studied intermediate redshift regime and the highest redshift data point at $z=1.71$. We present deep, joint \textit{XMM-Newton} and \textit{Chandra} X-ray observations for 10 of the most massive known galaxy clusters at redshifts $1.05 < z < 1.71$. 
For each cluster, we measure the metallicity in two spatial regions, an inner region ($r < 0.3r_{500}$) and an outer region ($r > 0.3r_{500}$)\footnote{$r_{500}$ is defined as the characteristic radius at which the total mass enclosed has a mean density 500 times the critical density of the Universe, $\rho_{\rm cr}(z)$.}. Our data quality is sufficient to provide interesting constraints in every case.

Our paper is organized as follows: in Section~\ref{sec:data}, we discuss the sample of clusters and describe the observations and reduction of the data. In Section~\ref{sec:methods}, we detail the method of analysis and the models used when fitting. In Section~\ref{sec:results}, we present the results of the analysis. In Section~\ref{sec:discuss}, we discuss the implications of our measurements in constraining  the evolution of metallicity at intermediate radii. We conclude in Section~\ref{sec:conc}.

In this paper, unless otherwise noted, all measurements are reported as the mode and associated 68.3\% credible interval corresponding to the highest posterior probability density.  We assume a flat $\Lambda$CDM cosmology with parameters $H_0 = 70 $ km\,s$^{-1}$\,Mpc$^{-1}$, $\Omega_\mathrm{m} = 0.3$, and $\Omega_{\Lambda} = 0.7$.
Metallicities are reported relative to Solar abundance measurements of \cite{Asplund0909.0948}.

\section{Data Selection and Reduction}
\label{sec:data}

The 10 clusters studied here have all been observed by both \textit{Chandra} and \textit{XMM-Newton}. They can be divided into two subgroups, based on the selection criteria of the surveys that first identified them:
\begin{enumerate}
    \item Seven clusters were identified by their Sunyaev-Zel'dovich (SZ) effect signal as part of the 2500 deg$^2$ survey by the South Pole Telescope (SPT) collaboration \citep{Bleem1409.0850}. Our targets are the highest-redshift, most massive objects in that sample. While previously studied for their thermodynamic properties \citep{Ghirardini2004.04747}, the present work focuses on measuring the metallicity of the ICM.
    \item The remaining three objects were identified from the ROSAT Deep Cluster Survey (RDCS1252.9$-$2927; \citealt{Rosati0309546}), the \textit{XMM-Newton} Large Scale Structure Survey (XLSSJ022403.9$-$041328; \citealt{Maughan0709.2300}) and a serendipitous detection of an extended X-ray source within an archival \textit{XMM-Newton} observation (1WGA J2235.3$-$2557; \citealt{Mullis0503004}) included in the WGACAT catalog of ROSAT sources (\citealt{White+2000}).
\end{enumerate}

\begin{figure}
	\includegraphics[width=\columnwidth]{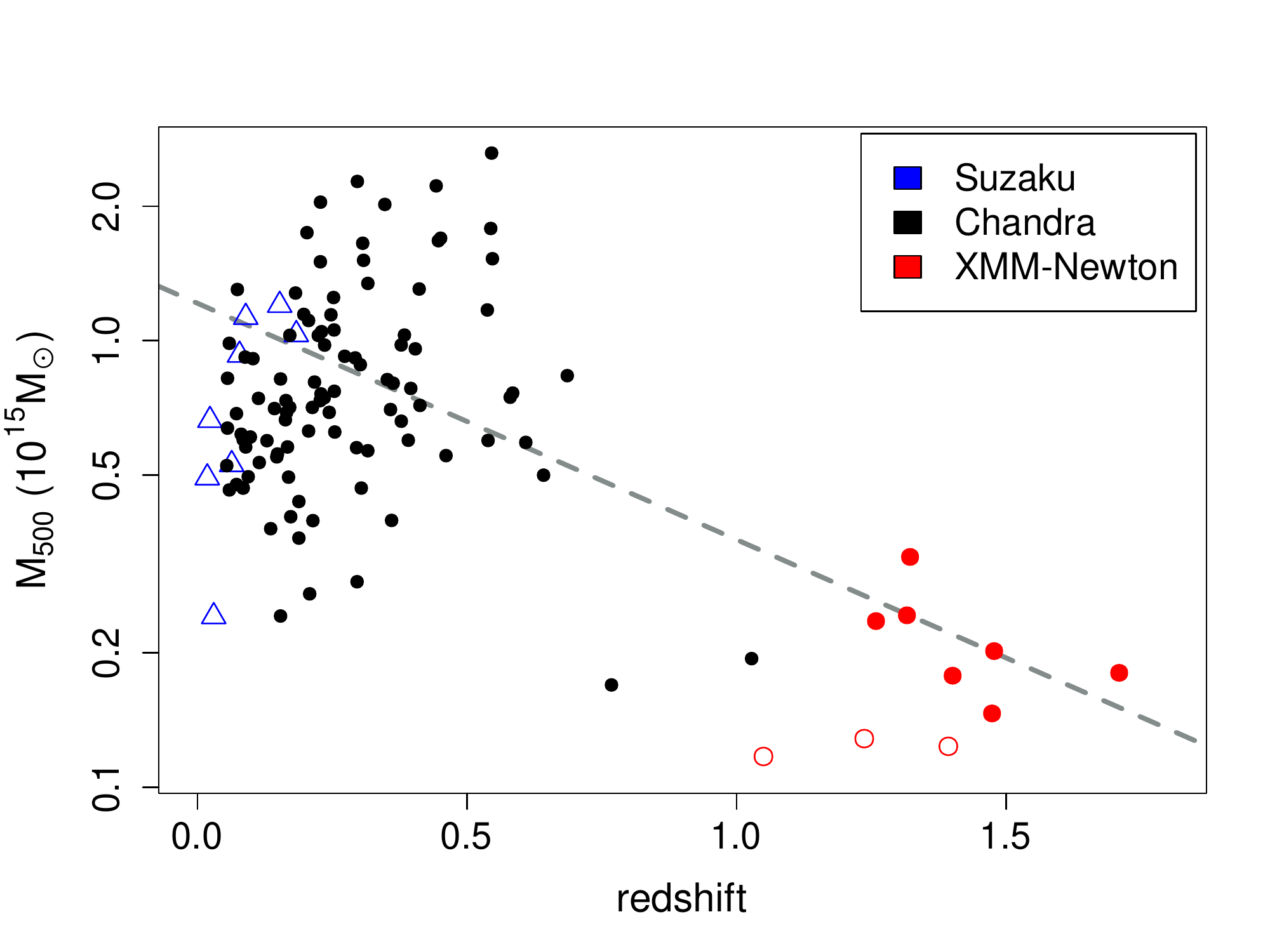}
    \caption{Mass-redshift diagram of the present sample, alongside clusters at low and intermediate redshifts that have previously been used to measure ICM metalicity outside cluster cores. Blue points denote low-redshift clusters previously studied using \textit{Suzaku} (metallicities and masses from \protect\citealt{Simionescu1302.4140, Werner1310.7948, Mantz1606.03407, Urban1706.01567, Mirakhor2007.13768}). Black points represent intermediate-redshift clusters analyzed by \protect\cite{Mantz1706.01476} using \textit{Chandra}. Red points show the clusters analyzed in this work. (Targets selected by their SZ-effect or extended X-ray emission are indicated by filled and open circles, respectively; see Section~\ref{sec:data}.) While their masses are smaller on average, the high-redshift clusters in our sample are representative of the progenitors of the low-redshift clusters, based on the expected growth rate of massive clusters (dashed curve; \protect\citealt{Fakhouri1001.2304}).}
    \label{fig:mass_redshift}
\end{figure}

Figure~\ref{fig:mass_redshift} shows the location of the ten clusters in our study in the mass-redshift plane, along with the systems at lower redshifts previously used by \cite{Mantz1706.01476} to study ICM metallicity evolution.  
All of the clusters in our study have redshifts measured with optical/IR spectroscopy (see Table \ref{tab:XMM}), except for SPT$-$CL J0459$-$4947; the redshift for this object was determined from spectral fits of the X-ray data (Section \ref{sec:spectra}). The OBSIDs for each cluster, along with the clean exposure times for \textit{XMM-Newton} and \textit{Chandra}, their J2000 coordinates, and redshifts are listed in Tables~\ref{tab:XMM} and~\ref{tab:chandra}. Hereafter, all clusters will be referred to by their survey/catalog designation (e.g. SPT, RDCS, 1WGA), followed by four digits corresponding to their Right Ascension, with the exception of XLSSJ022403.9$-$041328 (hereafter XLSSC 029).

\begin{table*}
	\centering
	\caption{Cluster names, redshifts, J2000 coordinates and \textit{XMM-Newton} OBSIDs for each cluster, together with the associated clean exposure times for each of the three detectors (MOS1, MOS2, pn). Redshifts are from (1) \protect\cite{Maughan0709.2300}, (2) \protect\cite{Rosati0309546}, (3) \protect\cite{Khullar1806.01962}, (4) \protect\cite{Stalder1205.6478}, (5) \protect\cite{Mullis0503004}, and (6) \protect\cite{Bayliss1307.2903}. SPT$-$CL J0459$-$4947 does not have an optical/IR spectroscopic redshift; the redshift was therefore fitted as a free parameter in the X-ray spectral analysis (\ref{sec:spectra}). 
	}
	\begin{tabular}{lccccccc}
		\hline
		Cluster & Redshift & R.A. (deg) & Dec. (deg) & XMM OBSID & MOS1 (ks) & MOS2 (ks) & pn (ks) \\
		\hline 
		XLSSJ022403.9$-$041328 & 1.05$^1$ & 36.0164 & -4.2248& 0210490101& 79& 80& 58\\ 
		\\
		RDCSJ1252.9$-$2927 & 1.237$^2$ & 193.2270 & -29.4548& 0057740301 & 47 & 48 & 39\\
		 &  &  & & 0057740401 & 64 & 64 & 55 \\ 
		\\
		SPT$-$CL J2341$-$5724 & 1.259$^3$ & 355.3513 & -57.4161& 0803050301& 93& 92& 77\\ 
		\\
		SPT$-$CL J0640$-$5113 & 1.316$^3$ & 100.0721& -51.2176& 0803050101& 103& 107& 79\\
		 &  &  & & 0803050701& 7& 7& 6 \\ 
		\\
		SPT$-$CL J0205$-$5829 & 1.322$^4$ &  31.4463& -58.4834& 0675010101& 23& 23& 19\\
		\\
		1WGAJ2235.3$-$2557 & 1.393$^5$ &  338.8359& -25.9618& 0311190101& 53& 53& 45\\
		\\
		SPT$-$CL J0607$-$4448 & 1.401$^3$ &  91.8958& -44.8039& 0803050501& 25& 25& 21\\
		 & &  & & 0803050801& 35& 35& 29\\ 
		\\
		SPT$-$CL J0313$-$5334 & 1.474$^3$ &  48.4813& -53.5744& 0803050401& 82& 83& 67\\
		 &  &  & & 0803050601& 66& 66& 51\\
		\\
		SPT$-$CL J2040$-$4451 & 1.478$^6$ &  310.2413& -44.8613& 0723290101& 75& 74& 62\\
		\\
		SPT$-$CL J0459$-$4947 & $1.71\pm0.02$ &  74.9227& -49.7823& 0801950101& 101& 103& 89\\
		 & &  & & 0801950201& 97& 96& 85\\
		 & &  & & 0801950301& 95& 95& 82\\
		 & &  & & 0801950401& 67& 66& 52\\
		 & &  & & 0801950501& 18& 19& 14\\
		\hline
	\end{tabular}
	\label{tab:XMM}
\end{table*}

\begin{table}
    \centering
    \caption{Cluster names, \textit{Chandra} OBSIDs and corresponding clean exposure times in either the ACIS-S or ACIS-I cameras.
    }
    \begin{tabular}{lccc}
        \hline
        Cluster & OBSID & Exposure (ks) & ACIS (S/I)\\
        \hline 
        XLSSJ022403.9$-$041328 & 6390 & 11 & S\\
        & 6394 & 16& S\\ 
		& 7182 & 22& S\\
		& 7183 & 19& S\\
		& 7184 & 23& S\\ 
		& 7185 & 32& S\\ 
		\\
		RDCSJ1252.9$-$2927 & 4198 & 163  & I\\ 
		& 4403 & 26& I\\ 
		\\
		SPT$-$CL J2341$-$5724 & 17208 & 54 & I\\ 
		& 18353& 44& I\\ 
		\\
		SPT$-$CL J0640$-$5113 & 17209 &  27& I\\ 
		& 17498 & 23& I \\ 
		& 18767 & 13& I\\
		& 18784 & 16& I\\ 
		\\
		SPT$-$CL J0205$-$5829 & 17482 & 50 & I \\ 
		\\
		1WGAJ2235.3$-$2557 & 6975 & 44 & S\\ 
		& 6976 & 24& S\\ 
		& 7367 & 80& S\\
		& 7368 & 33& S\\
		& 7404 & 15& S\\ 
		\\
		SPT$-$CL J0607$-$4448 & 17210 & 34& I \\ 
		& 17499 & 36& I\\
		& 17500 & 16& I\\
		& 18770 & 15& I\\ 
		\\
		SPT$-$CL J0313$-$5334 & 17212 &  22& I\\
		& 17503 & 38& I\\
		& 17504 & 21& I\\
		& 18847 & 21& I\\ 
		\\
		SPT$-$CL J2040$-$4451 & 17480 &  87& I\\ 
		\\
		SPT$-$CL J0459$-$4947 & 17211 &  13& I\\
		& 17501 &  22& I\\
		& 17502 &  14& I\\
		& 18711 &  23& I\\
		& 18824 &  22& I\\
		& 18853 &  30& I\\
		\hline
    \end{tabular}
    \label{tab:chandra}
\end{table}

\subsection{XMM-Newton}
\label{sec:XMM}
The data for each \textit{XMM-Newton} observation were reduced following the \textit{XMM-Newton} Extended Source Analysis Software ({\sc xmm-esas}; version 18.0.0)\footnote{\href{https://www.cosmos.esa.int/web/xmm-newton/sas}{https://www.cosmos.esa.int/web/xmm-newton/sas}} based on \cite{Snowden0710.2241} and the guidance from Snowden \& Kuntz in the {\sc xmm-esas} cookbook\footnote{\href{https://heasarc.gsfc.nasa.gov/docs/xmm/esas/cookbook/}{https://heasarc.gsfc.nasa.gov/docs/xmm/esas/cookbook/}}. Basic reduction was performed on both types of EPIC detectors (MOS and pn) using the {\sc emchain} and {\sc epchain}, and {\sc mos-filter} and {\sc pn-filter} tools, in order to generate cleaned event files and remove periods with heightened X-ray background. Each event file was also inspected manually to remove any background flares missed by the automatic process. Final clean exposure times for the MOS and pn detectors are given in Table~\ref{tab:XMM}. 
We also used standard ESAS tools to extract exposure maps, non-X-ray background maps, and images in the 0.4-4.0 keV energy band (observer frame), and to extract spectra, response matrices, and ancillary response files in various spatial regions for our spectral analysis. The use of these maps and spectra is detailed in Section~\ref{sec:methods}. In total, the clean exposure times for each of the three \textit{XMM-Newton} cameras are 1.13 Ms (MOS1), 1.14 Ms (MOS2), and 0.93 Ms (pn).

\subsection{Chandra}
\label{sec:Chandra}
All clusters in our sample also have \textit{Chandra} observations available on the Chandra Data Archive (CDA)\footnote{\href{https://cxc.harvard.edu/cda/}{https://cxc.harvard.edu/cda/}}. These data were reprocessed in the same manner as \citet{Mantz1502.06020} using version 4.6 of the \textit{Chandra} software analysis package, {\sc ciao}\footnote{\href{http://cxc.harvard.edu/ciao/}{http://cxc.harvard.edu/ciao/}} and version 4.71 of the \textit{Chandra} Calibration Database ({\sc caldb\footnote{\href{https://cxc.harvard.edu/caldb/}{https://cxc.harvard.edu/caldb/}}}). Second level event files were obtained for each cluster and the data were filtered to remove periods of high background during each observation. The \textit{Chandra} blank-sky data\footnote{\href{https://cxc.cfa.harvard.edu/ciao/threads/acisbackground/}{https://cxc.cfa.harvard.edu/ciao/threads/acisbackground/}} was used to generate quiescent background maps for each observation which were rescaled using measured count rates in the 9.5-12.0 keV range. We generated images, sky backgrounds and exposure maps in the 0.6-2.0 keV energy band (observer frame). These were used to determine cluster centers following the procedure of \cite{Mantz1502.06020}, to identify point source contaminants in the field of view, and to characterize the surface brightness profiles for each cluster. (See Sections~\ref{sec:pntsrc},~\ref{sec:psf}~and~\ref{sec:spectra}). In general, the \textit{Chandra} data are too shallow to be useful for the metallicity analysis, although a single spectrum was extracted to model a particularly strong AGN (See Section~\ref{sec:spectra}). \textit{Chandra} OBSIDs and ACIS-S/ACIS-I exposure times are given in Table~\ref{tab:chandra}. The final combined exposure for all \textit{Chandra} observations was 1.15 Ms.

\section{Methods and Modeling}
\label{sec:methods}

\subsection{Point Sources}
\label{sec:pntsrc}
For our purposes, the narrow Point Spread Function (PSF) of \textit{Chandra} allows for the identification and removal of point source contaminants such as AGN. Our process to account for these objects (as well as the processes described in later subsections) largely follows that of \cite{Mantz2006.02009}. Where available, we have used the identifications and measurements of point sources/AGN in the \textit{Chandra} cluster fields available from the Cluster AGN Topography Survey (CATS; Canning et al., in prep). Masks for these sources were applied to the \textit{XMM-Newton} observations, accounting for the AGN fluxes and \textit{XMM-Newton} PSF. In addition, the \textit{XMM-Newton} images were checked for additional sources not detected in the \textit{Chandra} data; masks for these sources were applied manually. An example of this process, showing the point sources identified in \textit{Chandra} images and masks applied to the XMM data can be found in Figure~\ref{fig:example_images}. The stacked XMM images for all clusters in this sample can be found in Appendix~\ref{sec:images}. \textit{}For the clusters in our sample for which CATS results were not available, masks were created by examining the second level \textit{Chandra} event file, then resized after superimposing them on stacked \textit{XMM-Newton} images. In two cases a point source could not be excised without removing significant cluster signal. In these cases, flux from the AGN was forward-modeled into our spectral analysis (Section~\ref{sec:spectra}).

\begin{figure*}
	\includegraphics[width=\textwidth]{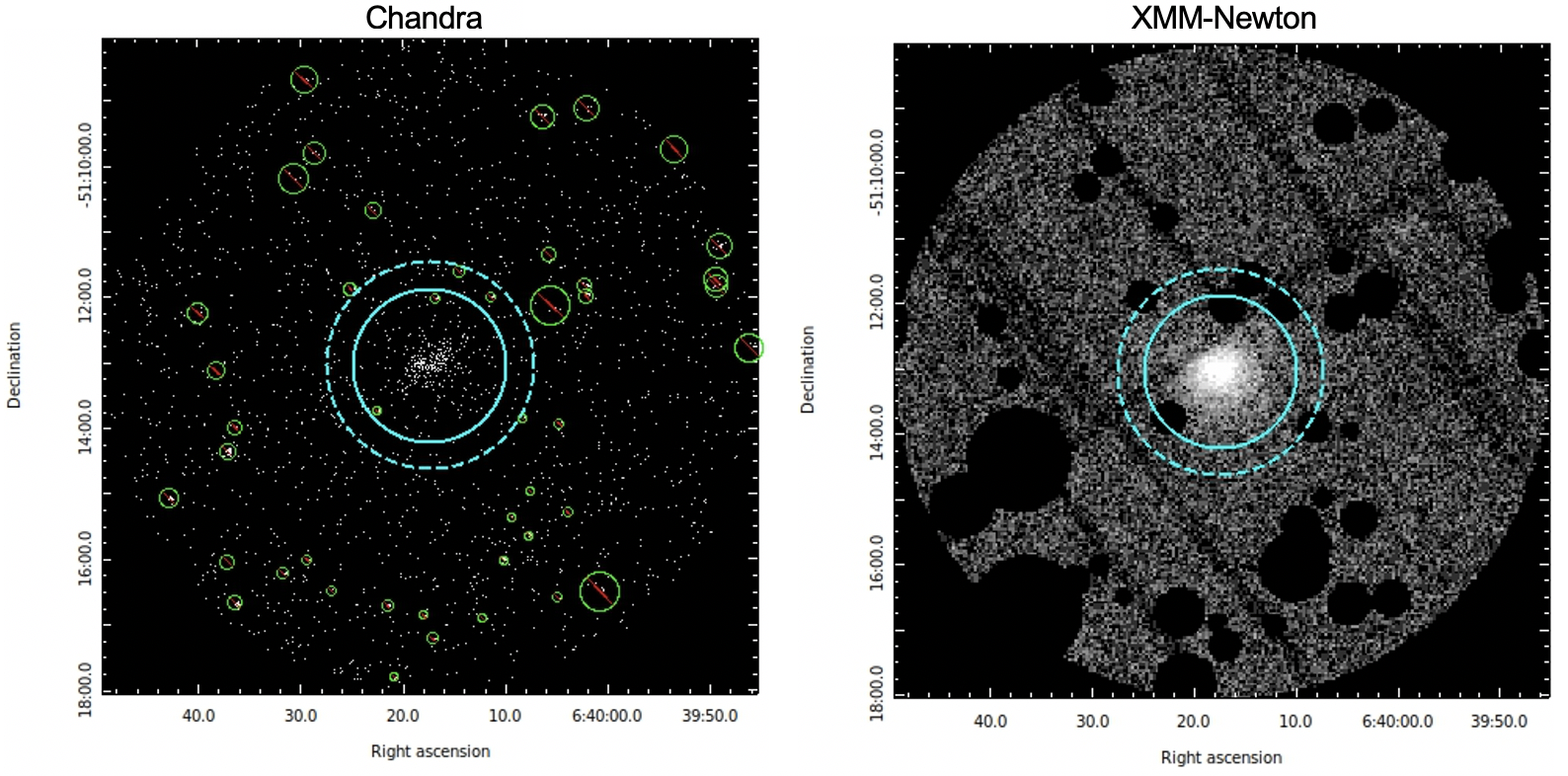}
    \caption{Left: Stacked \textit{Chandra} image for SPT~J0640. The point sources identified (green circles) are excluded in the analysis of the \textit{XMM-Newton} data. Cyan circles denote radii of $r_{500}$ (solid) and the visible extent of cluster emission from XMM images (dashed; see Section~\ref{sec:psf}).
    Right: Stacked \textit{XMM-Newton} image for SPT~J0640. The masked circular regions corresponding to point sources identified in the Chandra data, scaled according to the source flux and size of the XMM PSF, are clearly evident}
    \label{fig:example_images}
\end{figure*}

\subsection{Surface Brightness and PSF}
\label{sec:psf}
The relatively broad \textit{XMM-Newton} PSF must be accounted for in the analysis. We model this following \cite{Read1108.4835} as the sum of an extended $\beta$-profile and Gaussian core. To validate our PSF modeling, we compared fits to the surface brightness (SB) profiles for the clusters measured separately by \textit{XMM-Newton} and \textit{Chandra}. We extracted SB profiles from the masked images for each instrument (see sections~\ref{sec:data},~\ref{sec:pntsrc}) and converted both to consistent intensity units using the { \sc pimms}\footnote{\href{https://heasarc.gsfc.nasa.gov/cgi-bin/Tools/w3pimms/w3pimms.pl}{https://heasarc.gsfc.nasa.gov/cgi-bin/Tools/w3pimms/w3pimms.pl}}$^,$\footnote{\href{https://cxc.harvard.edu/toolkit/pimms.jsp}{https://cxc.harvard.edu/toolkit/pimms.jsp}} tool, assuming a metallicity of 0.3 solar and previously reported redshifts (Table~\ref{tab:XMM}) and temperatures. Where no temperature was available, we assumed a fiducial value of 6 keV.
We then fitted the two surface brightness profiles by the sum of a $\beta$-model and fitted background component:

\begin{equation}
S(r) = S_0\left[1+\left(\frac{r}{r_{\mathrm{c}}}\right)^2\right]^{-3\beta+0.5} + b,
\label{eqn:beta}
\end{equation}
characterized by a normalization ($S_{\mathrm{0}}$), core radius ($r_{\mathrm{c}}$), power law slope ($\beta$), and background ($b$).

\begin{figure*}
	\includegraphics[width=\textwidth]{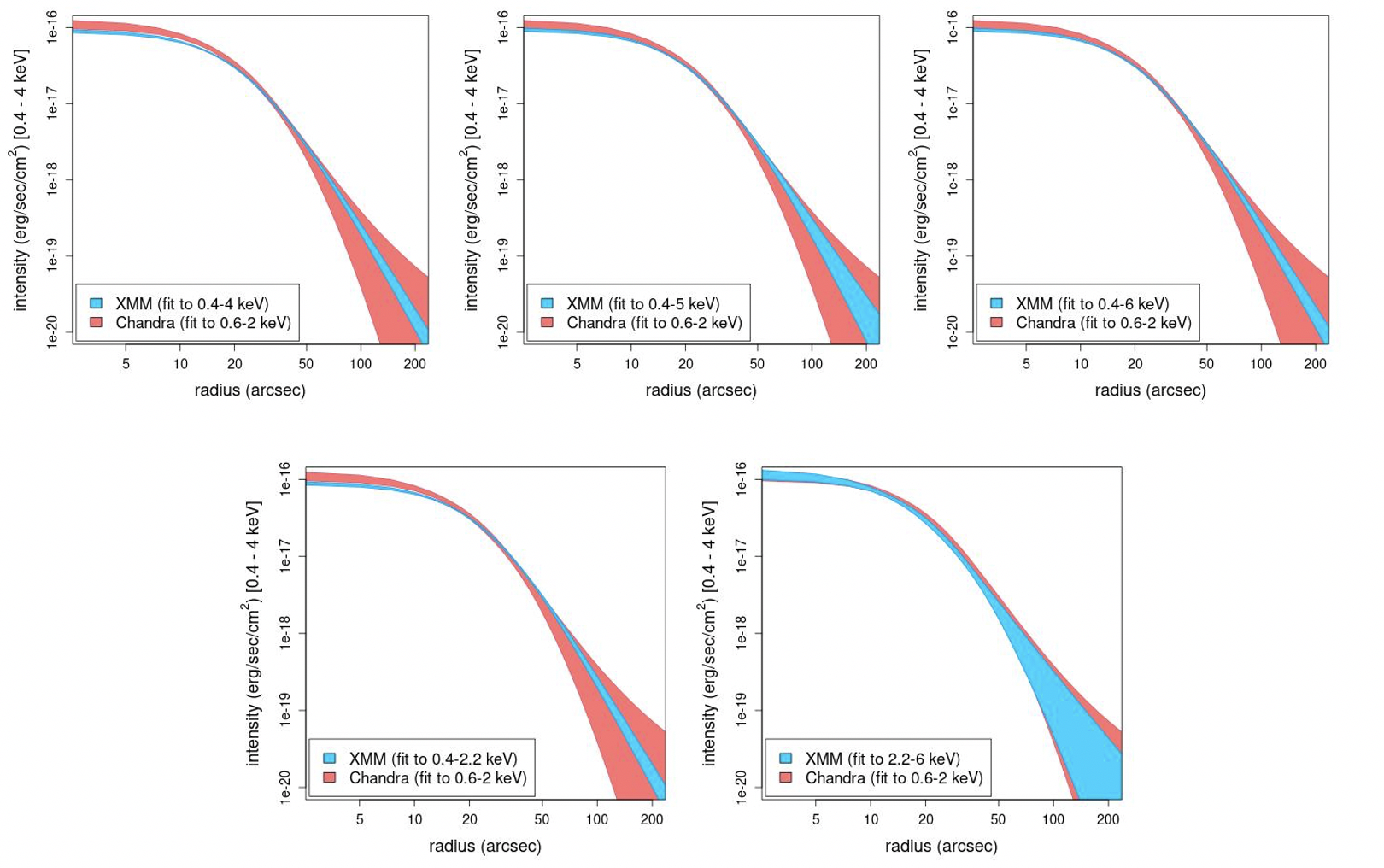}
    \caption{Comparison of independent $\beta$-model fits to both \textit{XMM-Newton} and \textit{Chandra} surface brightness profiles (SBP) for SPT~J0640. Counts from each detector have been converted to the same intensity units using {\sc pimms}. The shaded regions are the 68.3 per cent confidence constraints. The same \textit{Chandra} fit is used throughout (0.6$-$2.0 keV), while fits were made to XMM SBPs extracted from images generated in 5 different energy bands. The agreement observed validates the \textit{XMM-Newton} PSF model used in the analysis.
    }
    \label{fig:multiband_comp}
\end{figure*}

For \textit{XMM-Newton}, 
we convert the $\beta$-model in Equation~\ref{eqn:beta} from intensity units to counts, convolve it with the model for the \textit{XMM-Newton} PSF and compare the model to the measured counts via the Cash statistic \citep{Cash1979ApJ...228..939}. For \textit{Chandra}, the sharp instrumental PSF can be neglected in the $\beta$-model fit. Constraints on the model parameters are obtained by using the RGW\footnote{\href{https://github.com/abmantz/rgw}{https://github.com/abmantz/rgw}} implementation of Markov Chain Monte Carlo (MCMC) methods. 
Following \citet{Mantz2006.02009}, we check the consistency of the models fitted independently to the Chandra and XMM data, finding good agreement in all cases. Figure~\ref{fig:multiband_comp} shows that this agreement holds when several different XMM energy bands are used, indicating that our implementation of the PSF model, and in particular the assumption that it is constant with energy, is sufficient for our purposes. The agreement with \textit{Chandra} and the shape of the profile did not significantly change using our final measured temperatures. 
During this process, we also took note of the radius at which the cumulative enclosed counts as a function of radius flatten, i.e. the radius at which the cluster emission becomes negligible compared with the background. On average, this radius is $\sim2'$ and determines the outermost distance for extracting spectra (Section~\ref{sec:spectra}).

\subsection{Spectral Analysis}
\label{sec:spectra}
To model the \textit{XMM-Newton} spectral data, we use the {\sc xspec}\footnote{\href{https://heasarc.gsfc.nasa.gov/docs/xanadu/xspec/}{https://heasarc.gsfc.nasa.gov/docs/xanadu/xspec/}} analysis package (version 12.10.1s). Specifically, we seek to model the Bremsstrahlung continuum and emission from the iron line complex around 6.7 keV (rest frame) in the ICM. To do this we use {\sc apec} plasma models (\citealt{Smith0106478}; ATOMDB version 3.0.9) in which the emission is parametrized by a single temperature, density and metallicity. The redshifts of the clusters are fixed at the values determined from optical spectroscopy. We account for photoelectric absorption from gas in our own galaxy using the multiplicative {\sc phabs} model, fixed to the appropriate equivalent hydrogen column density \citep{HI4PI1610.06175} and using the cross sections of \cite{Balucinska1992ApJ...400..699B}.

In our models, we must also account for the fact that the two-dimensional spectral image of a galaxy cluster is a projection of its three-dimensional emission. To determine the intrinsic properties of our clusters, we first separate their emission into concentric annuli, centered on the coordinates listed in Table~\ref{tab:XMM}. Under the assumption of spherical symmetry, the annular data can then be modeled to infer the properties of the corresponding three-dimensional spherical shells, recognizing that part of the emission originating from a given spherical shell is projected onto all interior annuli \citep{Fabian1981ApJ...248...47F, Kriss1983ApJ...272..439K}. In our case, we must also account for the effects of the \textit{XMM-Newton} PSF (Section~\ref{sec:psf}), which redistributes counts in the two dimensional spectral image, as well as the area missing from annuli due to the AGN masks. These effects can be combined into a square mixing matrix that describes how much of the emission detected in a given annulus originates from each spherical shell. 

This mixing matrix is used to link normalizations of the {\sc apec} models fitted to the annuli appropriately for the deprojection analysis. The temperatures and metallicities of neighboring shells are linked over certain scales, depending on the specific analysis (i.e.\ density deprojection vs. temperature and metallicity profiles).

The spectral data were fitted over the 0.5-7.0 keV energy band (observer frame) with the following exclusions: 

\begin{enumerate}
    \item 1.2 - 1.65 keV band in the EPIC pn detector contaminated by Al instrumental emission lines
    \item 1.2 - 1.9 keV band in the EPIC MOS detectors due to contamination by Al and Si instrumental emission lines
    \item 4.4 - 5.7 keV band in all \textit{XMM-Newton} EPIC detectors due to contamination from Ti+V+Cr X-ray fluorescence lines.
\end{enumerate}

The instrumental and particle backgrounds for each cluster are modelled using spectra extracted from $3'$$-$$5'$ annuli, located well beyond the visible extents of the clusters. In the rare event that the extent of visible cluster emission extends beyond 2' from the cluster center, the innermost edge of the background region was extended to leave a $1'$ gap from the outer edge of the detected emission. As noted in Section~\ref{sec:pntsrc}, there were two cases where emission from AGN within the cluster fields needed to be forward-modeled in our analysis. In these cases, a power law model characterized by photon index and normalization was added to the {\sc apec} models. For SPT~J0459, we follow the same approach as \cite{Mantz2006.02009}, assuming a power-law spectrum with a canonical photon index for unresolved AGN of $1.4$ to model the relatively faint AGN, with the expectation that a change or omission of this component would negligibly affect our results  \citep{Mantz2006.02009}. In the case of SPT~J0205, a spectrum was extracted from the \textit{Chandra} data in a circular region of radius 2.5" centered on the AGN. A background spectrum was also extracted from \textit{Chandra} data in a 4"$-$9" annulus surrounding the AGN and used to constrain the AGN model in parallel with the XMM cluster and AGN fits.

We constrained the model parameters using  and the {\sc lmc}\footnote{\href{https://github.com/abmantz/lmc}{https://github.com/abmantz/lmc}} MCMC code, with the likelihood being given by the modified C-statistic in {\sc xspec} \citep{Arnaud1996ASPC..101...17A} to account for the Poisson nature of the counts from the cluster and background regions, after binning the data to include at least one count per channel to eliminate bias from empty channels.\footnote{\href{https://heasarc.gsfc.nasa.gov/xanadu/xspec/manual/XSappendixStatistics.html}{https://heasarc.gsfc.nasa.gov/xanadu/xspec/manual/XSappendixStatistics.html}}$^,$\footnote{\citet[][Appendix A]{Mantz1706.01476} discuss the performance of the modified C-statistic for measurements of metallicity, in particular its lack of bias even in the low signal-to-background regime where measurements are consistent with a prior boundary at $Z=0$.} The single Chandra AGN spectrum for SPT~J0205 is simultaneously fit in the 0.6--7.0 keV band, yielding a photon index for the identified point source of $2.3\pm0.2$. In addition to the Fe K-$\alpha$ line complex, we explored whether Fe L-shell emission was detected from the clusters, which could in principle complicate the analysis \citep{Ghizzardi2007.01084}. For most of our sample, the high redshift caused these emission lines to shift below the 0.5$-$7.0 keV analysis band. For the lower redshift clusters in our sample ($z \lesssim 1.3$), we detected no evidence of Fe L-shell emission contaminating our results.

\section{Results}
\label{sec:results}

\subsection{Gas Density and Gas Mass Profiles}
\label{sec:density}
Our \textit{XMM-Newton} observations allow us to determine the density and mass profiles of the ICM with a resolution of $5''-10''$, well matched to the size of the instrumental PSF. The clusters were divided into 10-15 annular regions, depending on the visible extent of the emission (typically $> 1.5r_{500}$). Using the methodology outlined in Section~\ref{sec:methods}, we first modelled the emission from all spherical shells with a common temperature and metallicity, but independent normalizations. The normalization acts as a proxy for emissivity, which can be converted into physical gas density, assuming a mean molecular mass of $\mu = 0.61m_p$, a reference cosmology, and the measured cluster redshift. These density profiles are integrated to determine the cumulative gas mass profiles for the clusters.

In order to compare and combine measurements of our target clusters, we need to define an appropriate reference radius, for which we adopt $r_{500}$ (typically about half of the virial radius). We compute the value of $r_{500}$ by solving the implicit equation
\begin{equation}
    M_{500} = M(r_{500}) = \frac{M_{\rm gas}(r_{500})}{f_{\rm gas}(r_{500})} = \frac{4}{3}\pi500\rho_{\rm cr}(z)r_{500}^3,
\end{equation}
where $\rho_{\rm cr}$ is the critical density of the Universe at the redshift of the cluster and $f_{\rm gas}(r_{500})$ is taken to be 0.125, based on X-ray measurements of clusters at redshifts of $z \lesssim 1.0$; \citep{Mantz1606.03407}. We expect this assumption to hold for the higher redshifts included in our sample, as the gas mass fraction likely does not evolve at intermediate radii for massive clusters (see \citealt{Eke9708070,Nagai0609247,Battaglia1209.4082,Planelles1209.5058,Barnes1607.04569,Singh1911.05751}). The values of $r_{500}$ for each cluster are reported in Table~\ref{tab:results}. We note that the results presented for ICM metallicity at intermediate radii in Section~\ref{sec:metals} are relatively insensitive to the precise value of $r_{500}$ and, thus, on the method used to estimate it.

\subsection{Metallicity Profiles}
\label{sec:metals}
Our main goal is to measure the metallicity of the ICM at intermediate radii. However, due to the size of the XMM PSF, we must simultaneously model emission from the gas both interior and exterior to this spatial region to obtain accurate results. Our X-ray data have sufficient spatial resolution and depth to measure the metallicity of each cluster in two independent bins: $0$$-$$0.3r_{500}$, $0.3r_{500}$$-$$r_{500}$. Beyond $r_{500}$, while the metallicity (and temperature) of the ICM cannot be measured precisely, the emissivity can still be determined, out to the maximum extent of the visible emission (Section~\ref{sec:psf}). In practice, we do this by binning the outermost spectrum into a single energy bin, thereby providing a measure of surface brightness in that region. The mixing matrix calculation accounts for the radial surface brightness distribution within each region and provides the correct links between models in our spectral fits. The inclusion of the emission from the outer regions in the analysis aids in the determination of robust deprojected results for the $0$$-$$0.3r_{500}$ and $0.3r_{500}$$-$$r_{500}$ shells. For this deprojection analysis, we assume that the emission from radii beyond $r_{500}$ has the same temperature and metallicity as the $0.3r_{500}$$-$$r_{500}$ shell. After obtaining initial estimates of the best-fit parameters by minimization of the C-stat statistic in XSPEC, we generated posterior distributions for our cluster model parameters via MCMC. To ensure that the temperature distributions reflect physically reasonable scenarios, we impose a prior on the temperature such that $kT<15$ keV. The metallicities measured in the inner and outer regions are reported in Table~\ref{tab:results}. We note that a simplified analysis of the {\it projected} spectra at intermediate radii (also accounting for PSF mixing between shells) returned consistent metallicity results. Note also that the fitted energy ranges and spatial regions described above differ somewhat from those used by \citet{Mantz2006.02009}, resulting in a modified value of core-excised ICM metallicity for SPT~J0459.

\begin{figure}
	\includegraphics[width=\columnwidth]{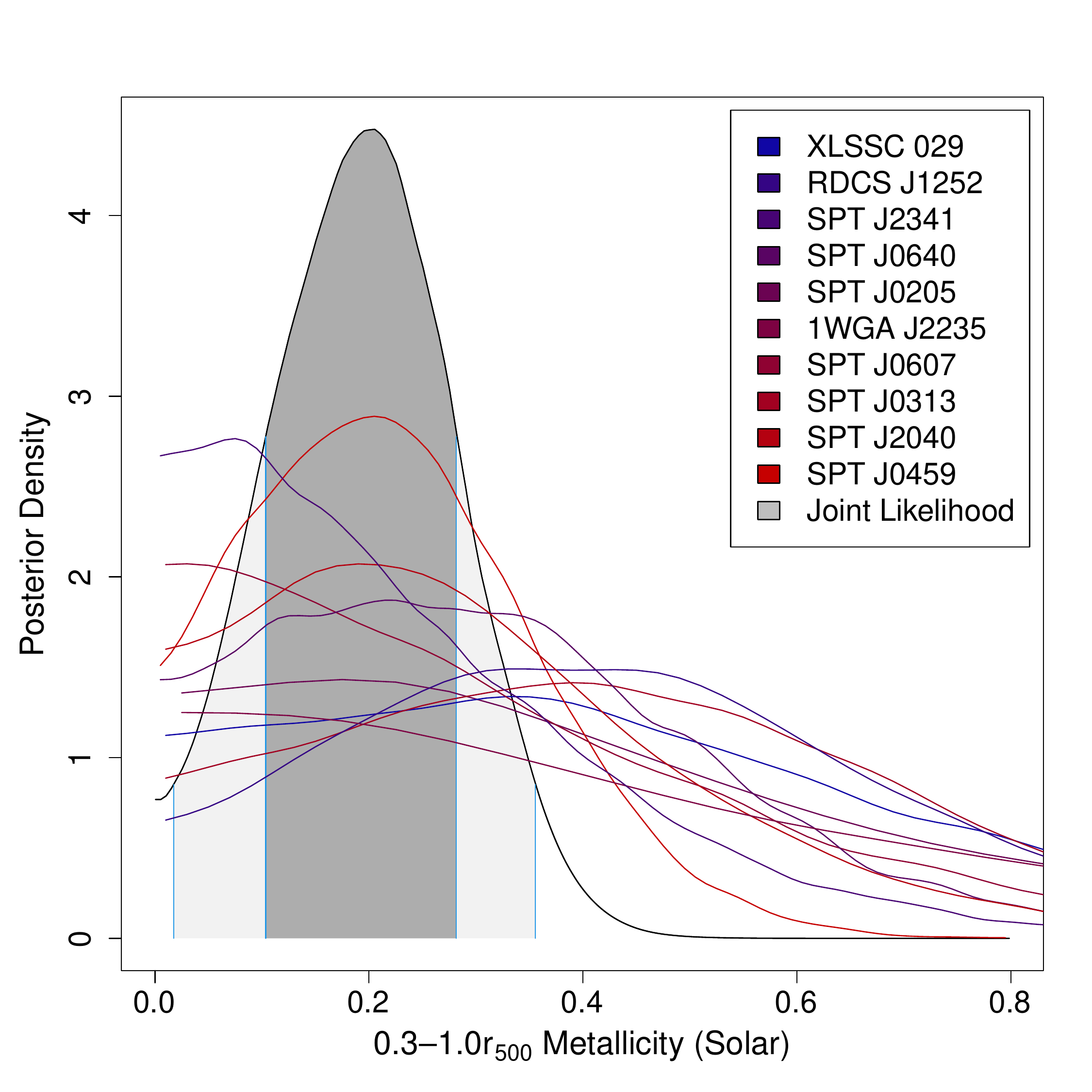}
    \caption{MCMC posteriors for the outer metallicity of each cluster, colored (blue $\rightarrow$ red) for increasing redshift. The shaded region corresponds to the combined posterior likelihood where a single true outer metallicity for all clusters is assumed. Dark and light shading, delineated by vertical blue lines, represent the 68.3\% and 95.4\% credible intervals of this combined distribution surrounding the most probable value of $Z/Z_{\odot}=0.21\pm0.09$.
    }
    \label{fig:combined_posterior}
\end{figure}

Figure~\ref{fig:combined_posterior} shows the posterior distributions of metallicity at intermediate radii in each cluster. Multiplying the individual posteriors yields a joint posterior PDF, assuming a common metallicity for all of the clusters. This resulting distribution of outer metallicity is also shown in Figure~\ref{fig:combined_posterior}, and yields a combined outer metallicity of $Z/Z_{\odot}=0.21\pm0.09$. Note that excluding SPT~J0459, which accounts for $\sim$30\% of the total XMM exposure time, yields a nearly identical constraint of $0.20\pm0.10$.

\begin{table*}
	\centering
	\caption{Values of $r_{500}$ amd $M_{500}$ as well as inner and outer metallicity measurements for the 10 high redshift clusters in this sample.
	}
	\renewcommand{\arraystretch}{1.5}
	\begin{tabular}{lccccc} 
		\hline
		Cluster &$r_{500}$ (Mpc) & $M_{500}$ (10$^{14} M_{\odot}$) & $Z/Z_{\odot} (0 - 0.3 r_{500})$  & $Z/Z_{\odot} (0.3 - 1.0 r_{500})$ \\
		\hline
		XLSSC 029  & 0.50 $\pm$ 0.03 & 1.2 $\pm$ 0.2 & $0.58^{+0.55}_{-0.50}$ & $0.33^{+0.24}_{-0.32}$\\
		RDCS~J1252  & 0.48 $\pm$ 0.03& 1.3 $\pm$ 0.3 & $0.38^{+0.80}_{-0.35}$ & $0.33^{+0.30}_{-0.20}$\\
		SPT~J2341  & 0.59 $\pm$ 0.03 & 2.4 $\pm$ 0.4 & $1.23 \pm 0.50$ & $0.08^{+0.22}_{-0.07}$ \\ 
		SPT~J0640  & 0.58 $\pm$ 0.03 & 2.4 $\pm$ 0.4 & $0.31^{+0.26}_{-0.20}$ & $0.22^{+0.20}_{-0.19}$\\
		SPT~J0205  & 0.64 $\pm$ 0.03 & 3.3 $\pm$ 0.5 & $0.00^{+0.65}_{-0.00}$ & $0.18^{+0.35}_{-0.15}$\\ 
		1WGA J2235 & 0.45 $\pm$ 0.02 & 1.2 $\pm$ 0.2 & $0.18^{+0.55}_{-0.15}$ & $0.00^{+0.70}_{-0.00}$\\ 
		SPT~J0607 & 0.51 $\pm$ 0.04 & 1.8 $\pm$ 0.4 & $0.58^{+0.35}_{-0.40}$ & $0.03^{+0.38}_{-0.02}$ \\
		SPT~J0313 & 0.46 $\pm$ 0.02 & 1.5 $\pm$ 0.2 & $0.23^{+0.83}_{-0.20}$ & $0.39^{+0.24}_{-0.30}$\\
		SPT~J2040 & 0.51 $\pm$ 0.03 & 2.0 $\pm$ 0.3 & $0.28^{+1.25}_{-0.25}$ & $0.19 \pm 0.18$\\ 
		SPT~J0459  & 0.457 $\pm$ 0.018 & 1.8 $\pm$ 0.2 & $0.67^{+0.30}_{-0.24}$ & $0.21^{+0.11}_{-0.15}$\\
		\hline
	\end{tabular}
	\label{tab:results}
\end{table*}

\section{Discussion}
\label{sec:discuss}
Over the past decade, evidence has solidified in support of the idea that the bulk of the enrichment of the ICM with metals occurs at early times, prior to galaxy cluster formation. Here, the key measurements have been: the radially and azimuthally uniform distribution of metals in the ICM observed out to large radii in the nearest, brightest galaxy clusters with the Suzaku satellite (\citealt{Werner1310.7948}, \citealt{Simionescu1302.4140}, \citealt{Simionescu1506.06164}); the consistent values of these metallicity measurements from cluster to cluster \citep{Urban1706.01567}; and the non-detection of  evolution in the metallicity of the ICM, beyond the inner regions ($0.3<r/r_{500}<1.0) $ and out to redshifts $z \sim 1.2$, albeit with significant uncertainties at the highest redshifts \citep{Ettori1504.02107,McDonald1603.03035,Mantz1706.01476}. The question remains, however, exactly when this enrichment occurred, and the most direct way to constrain this is to extend the measurements of cluster metallicities out to higher redshifts.

\begin{figure}
	\includegraphics[width=\columnwidth]{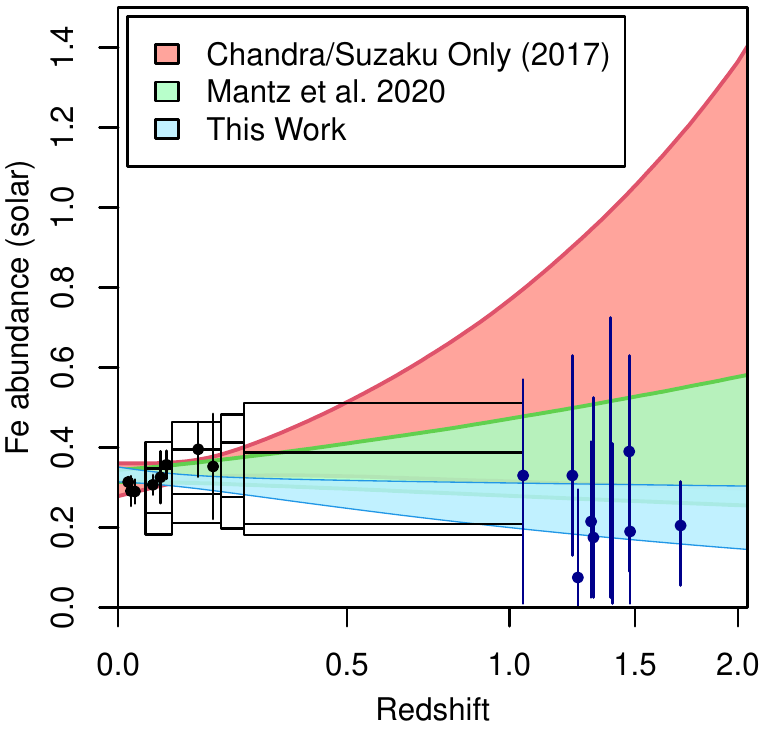}
    \caption[?]{68\% confidence constraints on the power-law evolution of outer metallicity as a function of redshift, $Z \propto (1+z)^{\gamma}$. Red constraints use observations from \textit{Suzaku} (black points) and \textit{Chandra} (binned black boxes; 68\% and 95\% confidence intervals; \citealt{Mantz1706.01476}). \textit{Suzaku} and \textit{Chandra} data have been rescaled to the \cite{Asplund0909.0948} Solar reference, and \textit{Chandra} measurements have an additional cross-calibration factor applied as described in \cite{Mantz1706.01476}. Green constraints add a measurement of SPT~J0459 at $z=1.71$ \citep{Mantz2006.02009}. 
    The blue region shows the new constraints from this work, with the inclusion of 10 deep cluster observations spanning $1.05 < z < 1.71$ (including a re-analysis of SPT~J0459). The new high-z data support a model in which significant enrichment of the ICM occurs at high redshifts ($z>2$), though they allow (and require at 68 per cent confidence) some ongoing enrichment at lower redshifts.}
    \label{fig:plaw_evo}
\end{figure}

\cite{Mantz2006.02009} presented results for SPT~J0459, the highest redshift cluster ($z = 1.71$) from which a measurement of a spatially resolved metallicity has been made to date. The present study has both re-analyzed those data and added results for a further nine clusters at $z>1$ (eight at $z>1.2$) to provide the most precise constraint on the metallicity of the ICM at intermediate radii and high redshift obtained to date.
The combined result for the $0.3r_{500}$$-$$r_{500}$ shell for these 10 clusters is $Z/Z_{\odot}=0.21\pm0.09$. 
We can compare this value to the metallicities measured in the outer parts of the Perseus Cluster ($Z/Z_{\odot}=0.314\pm0.012$; \citealt{Werner1310.5450}), the Coma Cluster ($Z/Z_{\odot}=0.29\pm0.04$; \citealt{Simionescu1302.4140}) and ten other nearby, massive systems ($Z/Z_{\odot}=0.316\pm0.012$; \citealt{Urban1706.01567}) studied with \textit{Suzaku}. Figure~\ref{fig:plaw_evo} shows the results for the high-redshift clusters studied here, together with the aforementioned \textit{Suzaku} results (also including a recent measurement from \citealt{Mirakhor2007.13768}) and results at intermediate redshift from \textit{Chandra}; \citealt{Mantz1706.01476}). The new results at high redshift are consistent (at the $\sim 68$ per cent confidence level), though slightly lower than, the previously reported results at intermediate and low redshift.

Following \cite{Ettori1504.02107} and \cite{Mantz1706.01476, Mantz2006.02009}, we have fitted the data in Figure~\ref{fig:plaw_evo} with a power law model for the evolution of ICM metallicity at intermediate radii as a function of redshift. The model fit is 
\begin{equation}
Z = Z_0\left(\frac{1+z}{1+z_\mathrm{piv}}\right)^{\gamma}.
\end{equation}
in which the pivot redshift $z_\mathrm{piv}$ is calculated to minimize the correlation of $Z_0$ with the power law slope $\gamma$. Additionally, we fit for a lognormal intrinsic scatter, $\sigma_{\ln Z}$, as well as a cross-calibration factor $\ln (Z/Z^\mathrm{Cha})$ necessary to renormalize \textit{Chandra} metallicity measurements as described in \cite{Mantz1706.01476}\footnote{While damage to the front illuminated \textit{Chandra} ACIS chips has generated charge transfer inefficiencies (CTI) that can bias \textit{Chandra} abundance measurements, there is currently no indication that \textit{XMM-Newton} results are affected in a similar manner. As such, we assume that results from Suzaku and XMM can be combined without the need for a cross-calibration factor in this analysis.}. The constraints on the evolution parameters of this model are detailed in Figure~\ref{fig:plaw_evo}. While previous analyses of low redshift data have provided excellent constraints on the "redshift zero" metallicity, the addition of these high redshift objects significantly improves the constraints on the evolution slope, $\gamma$. In combination with the lower-redshift data, we find $Z_0 = 0.321^{+0.014}_{-0.016}$ (at a redshift of $z_\mathrm{piv}=0.09$) and $\gamma = -0.5^{+0.4}_{-0.3}$. A summary of all fit parameters can be found in Table~\ref{tab:plaw_fit}. 

\begin{table}
    \centering
    \caption{Constraints on the model of power law evolution of ICM metallicity, $Z=Z_0\left(\frac{1+z}{1+z_\mathrm{piv}}\right)^{\gamma}$, as well as the intrinsic scatter and \textit{Chandra} cross-calibration factor. The constraint on metallicity is given for the pivot redshift that minimizes the correlation between the normalization and the evolution power law slope.
    }
    \begin{tabular}{lc}
        \hline
        $z_\mathrm{piv}$ & 0.09 \\
        \\
        $Z_0/Z_\odot$ & $0.321^{+0.014}_{-0.016}$ \\
        \\
        $\gamma$ & $-0.5^{+0.4}_{-0.3}$\\
        \\
        $\sigma_{\ln Z}$ & $< 0.09$ \\
        \\
        $\ln \left(Z/Z^\mathrm{Cha}\right)$ & $0.28^{+0.10}_{-0.07}$\\
        \hline 
    \end{tabular}
    \label{tab:plaw_fit}
\end{table}

Our results (see also \citealt{Mantz2006.02009}) provide the first precise measurement of the metallicity of ICM at high redshifts (z > 1) for the regions beyond cluster cores. The upper redshift limit of our sample corresponds to a lookback time of nearly 10 billion years, and our results have significant implications for models of ICM enrichment (e.g. \citealt{Biffi1701.08164, Biffi1801.05425, Vogelsberger1707.05318}). For the ICM in these high redshift systems to approach the levels of enrichment we see in local clusters today, a significant fraction of the production, and subsequent mixing, of these metals must have occurred at very early times, before the clusters formed and likely before redshift $z\sim 2$. Our results point to an intense early period of star formation and associated AGN activity in proto-cluster environments that both generated the metals and expelled them from their host galaxies into the surrounding intergalactic medium. These metals became well mixed within the intergalactic medium that later accreted onto clusters, providing a foundation for the near-uniform metallicity in cluster outskirts, both within individual clusters and from system to system, that we observe today. At the same time, our results provide a first tantalizing indication (albeit at $\sim 68$ per cent confidence) for a possible increase in the metallicity of the ICM at large radii from $\sim0.2\,Z_{\odot}$ at $z\sim2$ to $\sim0.3\,Z_{\odot}$ today. This late-time enrichment, if confirmed, must occur in a way that preserves the spatial uniformity of metal abundances seen in well studied, low-redshift clusters.

Our results also indicate the presence of central metallicity gradients (also at modest significance) in two of the clusters in our sample, SPT~J2341 and SPT~J0459 (see Table~\ref{tab:results}); the MCMC posteriors indicate $Z_{\mathrm{in}}/Z_{\mathrm{out}} \sim$2, with detections of central metallicity enhancements at the 96 and 90 per cent confidence level, respectively. Subcluster merger events are thought to be effective at disrupting central metallicity gradients (e.g. \citealt{Allen9802219,De-Grandi0012232,Rasia1509.04247}). The presence of metallicity gradients in these systems (at redshifts $z=1.26$ and $z=1.71$, respectively) may indicate that they have not yet undergone a merger event violent enough to disrupt and mix their central metallicity peaks.

While the tightening of the evolutionary model constraints with the addition of the data presented here is impressive, it should be noted that the investments of new \textit{Chandra} and \textit{XMM-Newton} observing time involved were substantial, with approximately 1Ms of clean exposure time provided by each telescope (i.e. Chandra and each of the three XMM-Newton telescopes) after event filtering. We additionally note that the clusters studied here include the most massive clusters at $z > 1$ discovered in the full SPT 2500 deg$^2$ survey \citep{Bleem1409.0850}. Future studies of this survey region will likely target predominantly less massive systems with lower emissivity. It may therefore be challenging to improve substantially on the measurements at high$-z$ presented here with existing technology and analysis methods. 

While measurements of low-redshift clusters with the \textit{Suzaku} satellite were able to probe the temperature and metallicity of the ICM out to radii well beyond $r_{500}$ and approaching the virial radius, observations with \textit{Chandra} and \textit{XMM-Newton} have to date been limited to $r<r_{500}$ by the instrumental background, sourced by the interactions of cosmic ray particles with the satellites and detectors. Our best near-term hope to improve substantially on measurements of the type presented here may therefore lie with approaches to reduce the impact of the particle background on such measurements \citep{Wilkins2012.01463}.

\section{Conclusions}
\label{sec:conc}
We have presented the analysis of deep \textit{Chandra} and \textit{XMM-Newton} observations of 10 massive, high redshift ($z>1$) galaxy clusters, selected from SZ and X-ray surveys, with the goal of obtaining improved constraints on the enrichment history of the intracluster medium. The X-ray data allow for the rigorous separation of emission from the ICM and contaminating point sources and robust estimates of $r_{500}$. For each cluster, we were able to measure the metallicity in two radial bins, spanning radii of ($0-0.3r_{500}$) and ($0.3-1r_{500}$). 
For the outer region, the combined measurement for all ten clusters, $Z/Z_{\odot} = 0.21 \pm 0.09$, is consistent with but slightly lower than the value of $\sim 0.3$ measured for low-redshift clusters. The data confirm that significant enrichment of the ICM occurs at very high redshifts ($z>2$), while leaving open the possibility that some enrichment at these radii continues at lower redshifts. Combining our results with previous measurements of lower redshift systems allows us to place the tightest constraints to date on models of the evolution of cluster metallicity at intermediate radii, yielding a power-law redshift evolution slope of $\gamma = 0.5^{+0.4}_{-0.3}$.

New observations of clusters at the same redshifts and imaging depth, utilizing similar technology and analysis techniques, will be challenged to improve significantly on the metallicity constraints presented here. In the near term, the development of novel methodologies for background reduction (e.g. \citealt{Wilkins2012.01463}) to improve the signal-to-noise of measurements from existing data should be pursued. This will hopefully provide our first access to information from beyond $r_{500}$ at intermediate and high redshifts. Looking further ahead, future flagship X-ray observatories such as ATHENA\footnote{\href{https://www.the-athena-x-ray-observatory.eu/}{https://www.the-athena-x-ray-observatory.eu/}} and \textit{Lynx}\footnote{\href{https://www.lynxobservatory.com/}{https://www.lynxobservatory.com/}} will allow us to observe and study $z>2$ clusters in detail, transforming our knowledge of the topics considered here.

\section*{Acknowledgements}

We acknowledge support from the National Aeronautics and Space Administration under Grant No.
80NSSC18K0578, issued through the XMM-Newton Guest
Observer Facility; and from the U.S. Department of Energy
under contract number DE-AC02-76SF00515.

This work was performed in the context of the South
Pole Telescope scientific program. SPT is supported by the
National Science Foundation through grant PLR-1248097.
Partial support is also provided by the NSF Physics Frontier
Center grant PHY-0114422 to the Kavli Institute of Cosmological Physics at the University of Chicago, the Kavli
Foundation and the Gordon and Betty Moore Foundation
grant GBMF 947 to the University of Chicago. The SPT is
also supported by the U.S. Department of Energy. Work
at Argonne National Lab is supported by UChicago Argonne LLC, Operator of Argonne National Laboratory (Argonne). Argonne, a U.S. Department of Energy Office of Science Laboratory, is operated under contract no. DE-AC02-
06CH11357

\section*{Data Availability}

The data underlying this article are available from the \textit{XMM-Newton} Science Archive (XSA) at \href{https://www.cosmos.esa.int/web/xmm-newton/xsa}{https://www.cosmos.esa.int/web/xmm-newton/xsa} as well as the \textit{Chandra} Data Archive (CDA) at \href{https://cxc.harvard.edu/cda/}{https://cxc.harvard.edu/cda/}. For searchable OBSIDs for each telescope, see Tables~\ref{tab:XMM} and~\ref{tab:chandra} respectively in this work.



\bibliographystyle{mnras}
\bibliography{high_z_metallicity} 




\appendix
\section{XMM-Newton Images}
\label{sec:images}

Presented here are stacked \textit{XMM-Newton} images in the 0.4-4 keV energy band (observer frame) for all 10 clusters used in this analysis.

\begin{figure*}
	\includegraphics[scale=0.9, trim={0 4.8cm 0 0},clip]{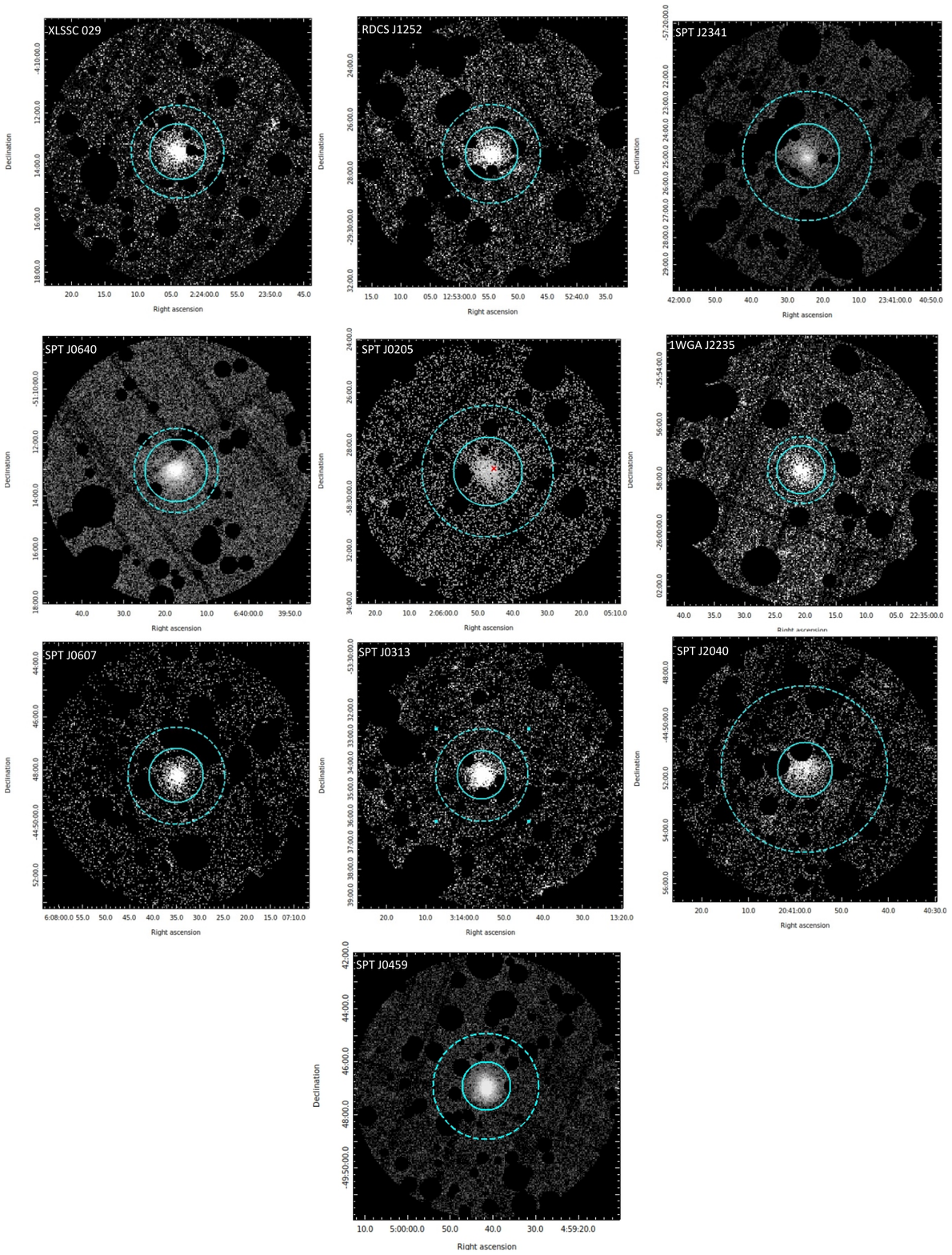}
    \caption{Stacked \textit{XMM-Newton} images for each of the 10 clusters analyzed in this work. Each image is generated from counts in the 0.4-4 keV energy band (observer frame) and is masked of point sources identified in the corresponding \textit{Chandra} imaging (see Section~\ref{sec:pntsrc}). Cyan regions denote the radial extent of $r_{500}$ (solid) and detectable cluster emission (dashed) for each cluster. The red cross in SPT~J0205 is placed at the position of the unmasked AGN forward-modeled in the spectral analysis.}
    \label{fig:XMM_images}
\end{figure*}





\bsp	
\label{lastpage}
\end{document}